\documentclass[a4paper]{article}

\usepackage{INTERSPEECH2022}

\title{Wider or Deeper Neural Network Architecture for Acoustic Scene Classification with Mismatched Recording Devices}
\name{Lam~Pham$^{1}$, 
        Khoa~Dinh$^{2}$,
        Dat~Ngo$^{3}$,
        Hieu~Tang$^{4}$, 
        Alexander~Schindler$^{1}$}     
\address{
  $^1$ Lam Pham and Alexander Schindler are with Austria Institute of Technology, Austria.\\
  $^2$ Khoa Dinh is with The University of Danang, Viet Nam. \\
  $^3$ Dat Ngo is with University of Essex, UK. \\
  $^4$ Hieu Tang is with Hongik University, Korea. 
  }
\email{Lam.Pham@ait.ac.at, tdkhoa@dut.udn.vn, thanhdat5494@gmail.com, tqhieu94@gmail.com, Alexander.Schindler@ait.ac.at}

\begin{document}

\maketitle
\begin{abstract}

In this paper, we present a robust and low complexity system for Acoustic Scene Classification (ASC), the task of identifying the scene of an audio recording.
We first construct an ASC baseline system in which a novel inception-residual-based network architecture is proposed to deal with the mismatched recording device issue.
To further improve the performance but still satisfy the low complexity model, we apply two techniques: ensemble of multiple spectrograms and channel reduction on the ASC baseline system. 
By conducting extensive experiments on the benchmark DCASE 2020 Task 1A Development dataset, we achieve the best model performing an accuracy of 69.9\% and a low complexity of 2.4M trainable parameters, which is competitive to the state-of-the-art ASC systems and potential for real-life applications on edge devices.
   
\end{abstract}

\section{Introduction}
\label{intro}
The Acoustic Scene Classification (ASC) task, one of main topics in `Machine Hearing' research field~\cite{lyon_machinehearing}, has attracted much research attention recently.
Indeed, not only more and more ASC datasets such as Litis Rouen~\cite{litis_dataset}, ESC50~\cite{esc50_data}, DCASE Task 1 challenges~\cite{dcase_com}, or Crowded Scenes~\cite{lam_crowdedscene_data} have been published, but various ASC systems, leveraging deep neural networks, have been also proposed (i.e. The literature review section in~\cite{lam_phd_thesis} summarises and analyses state-of-the-art ASC systems as well as updated machine learning and deep learning techniques applied for ASC).
Regarding ASC challenges, they mainly come from different noise resources, various sounds in real-world environments, occurring as
single sounds, continuous sounds or overlapping sounds, or dynamic energy of sound events in a sound scene recording. 
These challenges drive ASC research community to focus on analysing frequency bands~\cite{phaye_dca_18, interspeech_2021_01, damping} rather than specific sound events~\cite{hong_dca_18}.  
However, the new issue of mismatched recording devices firstly mentioned in DCASE 2018 Task 1B challenge~\cite{dcase_2018_1b} further increases ASC challenge as this issue causes energy distribution at certain frequency bands of spectrograms from the same class significantly different (i.e. In Figure 1 of~\cite{primus2019acoustic}, Mel-based spectrograms from the same sound scene of `on Tram' show different as they are from three different recording devices).  
To deal with the issue of mismatched recording device, ensembles of different spectrogram input~\cite{truc_dca_18_int, lam_dca_18, lam_j1_asc, ngo2020sound, lam_dca_16_int} or ensemble of multiple classification models~\cite{ens_01, ens_02} are mainly approached.
Indeed, almost ASC systems submitted to the recent DCASE 2020 Task 1A challenge~\cite{dcase_2020_1a} leverage ensemble-based models which prove robust for unseen samples as well as for dealing with the issue of mismatched recording devices.
However, ensemble methods present large footprint models, which is challenging to apply on edge devices or real-time applications.

In this paper, we aim to develop an ASC system which is robust to deal with ASC challenges mentioned recently.
To this end, we firstly construct a baseline system in which a novel neural network, a shallow and wide inception-residual-based architecture, is presented.
The proposed baseline is then compared with benchmark neural networks such as VGGish (e.g. VGG16, VGG19) or Residual (e.g. Resnet, DenseNet, MobileNet, or Xception) based architectures to evaluate whether a wider and shallow network or a deeper architecture is effective for ASC with mismatched recording device issue.
Then, we apply two techniques: (1) ensemble of multiple spectrograms and (2) channel reduction on the proposed baseline, achieve a low-complexity system but still perform robust for ASC.
\section{The proposed ASC baseline system} 
\label{baseline}
\begin{figure}[th]
    	\vspace{-0.3cm}
    \centering
    \includegraphics[width =1.0\linewidth]{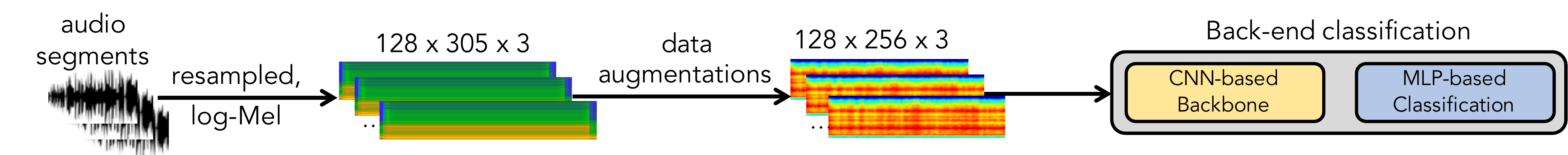}
   	\vspace{-0.3cm}
	\caption{High-level architecture of our ASC baseline system.}
    \label{fig:pp02}
\end{figure}
To evaluate our proposed systems and compare with the state-of-the-art ASC systems, we firstly present a baseline which shows a high-level architecture in Figure~\ref{fig:pp02}. 
As Figure~\ref{fig:pp02} shows, the proposed ASC baseline can be separated into three main steps: The front-end feature extraction, the online data augmentation, and the back-end classification.
\subsection{The front-end feature extraction}
\label{feature}
\begin{figure*}[t]
    \centering
    \includegraphics[width =1.0\linewidth]{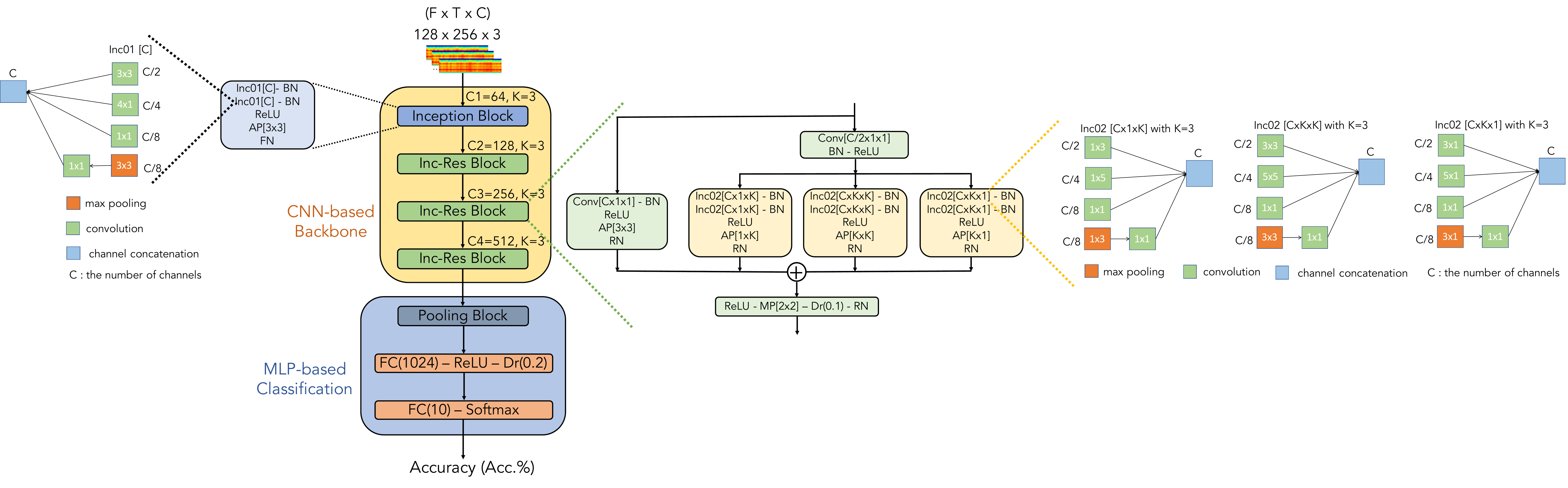}
	\caption{The proposed novel inception-residual-based deep neural network for the back-end classification.}
   	\vspace{-0.4cm}
    \label{fig:pp01}
\end{figure*}
%
The audio recordings are firstly resampled to 32,000 Hz. 
Then, they are transferred into log-Mel spectrogram using Librosa toolbox~\cite{librosa_tool}. 
By setting the Hanna window size, the hop size, the filter number to 2048, 1024, 128 and applying delta, delta-delta on each spectrogram, we generate a log-Mel spectrogram of 128$\times$305$\times$3 from one 10-second segment (Note the channel dimension is 3, which causes by concatenating the original log-Mel spectrogram, delta, and delta-delta).

\subsection{The online data augmentation}
\label{augmentation}

In this paper, we apply three data augmentation methods of Random Cropping~\cite{spec_crop}, Specaugment~\cite{spec_aug}, and Mixup~\cite{mixup1, mixup2}, respectively. 
In particular, the temporal dimension of log-Mel spectrograms of 128$\times$\textbf{305}$\times$3 is randomly cropped to 128$\times$\textbf{256}$\times$3 (e.g. Random Cropping method).
Then, ten continuous and random frequency or temporal bins of the cropped spectrograms are erased (e.g. Specaugment method).
Finally, the spectrograms are randomly mixed together using different ratios from Uniform or Beta distributions (e.g. Mixup method).
All of three data augmentation methods are applied on each batch of spectrograms during the training process, referred to as the online data augmentation.
\subsection{The back-end classification}
\label{network}
As Figure~\ref{fig:pp02} shows, the proposed back-end classification can be separated into two main parts: CNN-based deep neural network backbone and multiplayer perception (MLP) based classification.
In particular, the proposed CNN-based backbone comprises four blocks: one Inception Block and three Inc-Res Blocks as described at the upper part of Figure~\ref{fig:pp01}, which makes use of inception-based (e.g. Inception Block) or both inception-based and residual architectures (e.g. three Inc-Res Blocks).
Three Inc-Res Blocks share the same network architecture, but channel numbers increases from 128, 256, to 512, respectively. 
Four blocks of the CNN-based backbone are performed by Inception layers (Inc01[Channel] in Inception Block, Inc02[Channel$\times$Kernel Size] in Inc-Res Blocks), Convolutional layer (Conv[Channel$\times$Kernel Size]), Bach Normalization (BN)~\cite{batchnorm}, Dropout (Dr(Drop Ratio)~\cite{dropout}, Rectify Linear Unit (ReLU)~\cite{relu}, Max Pooling (MP[Kernel Size]), Average Pooling (AP[Kernel Size], Residual Normalization (RN($\lambda=0.4$)) inspired from~\cite{resnorm}).

Regarding two Inc01 layers used in Inception Block as shown in the left part of Figure~\ref{fig:pp01}, we use fixed kernel sizes of [3$\times$3], [1$\times$1], and [4$\times$1] (Note that using the kernel [4$\times$1] helps to focus on frequency bands).
Meanwhile, kernel sizes used in Inc02 layers in three Inc-Res Blocks as shown in the right part of Figure~\ref{fig:pp01} are defined by $K$ kernel size. 
By using different kernel sizes of [K$\times$1], [K$\times$K], and [1$\times$K], then applying AP layers with the same kernels, and finally adding output of these AP layers together, the network can learn distribution of energy in certain frequency bands effectively, which strengthens the network to deal with the issue of mismatched recording devices.

The MLP-based classification as shown in the lower part of Figure~\ref{fig:pp01} performs a Pooling Block and two fully connected layer blocks.
At the Pooling Block, we extract thee types of feature: (1) global average pooling across the channel dimension, (2) global max pooling across temporal dimension, and (3) global average pooling across frequency dimension.
We then concatenate these features before feeding into fully connected blocks.
While the first fully connected layer (FC[Channel]) combines with ReLU and Dr(0.2), the second fully connected layer uses Softmax layer for classifying into 10 scene categories.
\begin{table*}[t]
    \caption{Compare our proposed baseline (log-Mel-baseline) to benchmark networks on the DCASE 2020 Task 1A Development set across the different recording devices (Acc.\%)} 
        	\vspace{-0.2cm}
    \centering
    \scalebox{0.65}{

    \begin{tabular}{|c|c|  c|c|c|  c|c|c|  c|c|c| c|c|} 
        \hline 
	         &\textbf{Proposed}  &\textbf{VGG16}   &\textbf{VGG19}  &\textbf{MobileNetV1} &\textbf{MobileNetV2} &\textbf{ResNet50V2} &\textbf{ResNet101V2} &\textbf{ResNet152V2} &\textbf{DenseNet121} &\textbf{DenseNet169} &\textbf{DenseNet201} &\textbf{Xception} \\
	        &\textbf{Baseline}   & & & &   & & & &  & & & \\  
        \hline 
       
        \textbf{A(\%)}   &77.3  &68.3       &67.1 &74.2 &71.0   &74.1 &73.0 &74.0  &74.1 &75.7 &74.8  &75.2\\
        \textbf{B(\%)}   &72.0  &54.5       &56.1 &60.1 &56.3   &57.9 &57.8 &60.8  &63.1 &62.2 &58.3  &62.0\\
        \textbf{C(\%)}   &76.6  &61.5       &61.5 &63.7 &60.6   &63.7 &66.3 &67.8  &63.7 &68.6 &68.6  &68.4\\
        \textbf{S1(\%)}  &68.5  &55.3       &49.8 &57.2 &52.5   &60.2 &59.5 &52.5  &62.0 &57.2 &57.2  &60.1\\
        \textbf{S2(\%)}  &65.8  &54.4       &51.4 &51.4 &55.2   &54.1 &54.6 &52.8  &58.9 &55.7 &56.3  &54.7\\
        \textbf{S3(\%)}  &69.7  &53.5       &52.0 &55.4 &52.5   &55.6 &57.3 &57.0  &60.2 &61.0 &59.6  &62.2\\
        \textbf{S4(\%)}  &63.3  &43.8       &38.3 &43.8 &41.3   &45.6 &44.9 &47.6  &51.7 &50.1 &51.5  &50.4\\
        \textbf{S5(\%)}  &64.5  &45.3       &44.4 &44.7 &46.1   &52.0 &48.5 &44.3  &53.8 &52.2 &48.7  &49.4\\
        \textbf{S6(\%)}  &58.8  &40.1       &31.4 &32.6 &29.5   &31.7 &35.5 &31.9  &40.8 &43.7 &35.7  &35.2\\   
        
        \hline     
        \textbf{Average(\%)}  &68.5  &53.3  &50.8 &53.3 &51.6   &55.1 &55.2 &54.0   &58.7 &58.6 &56.7  &57.9\\         
        \hline 
        \textbf{Parameters(M)}  &9.6  &138.4  &143.7 &4.3 &3.5   &25.6 &44.7 &60.4   &8.1 &14.3 &20.2  &22.9\\         
        \hline 

    \end{tabular}
    }
    \label{table:res_01} 
\end{table*}
\begin{table}[t]
    \caption{Channel reduction to achieve low complexity models} 
        	\vspace{-0.2cm}
    \centering
    \scalebox{0.95}{

    \begin{tabular}{|l|c |c |c | c |c |} 
        \hline 
	           & \textbf{baseline} &\textbf{Red01}  &\textbf{Red02}  &\textbf{Red03}  \\
        \hline        
        \textbf{Inception Block} &2$\times$64  &64  &32  &16  \\
        \textbf{Inc-Res Block}   &2$\times$128 &128 &64  &32  \\
        \textbf{Inc-Res Block}   &2$\times$256 &256 &128 &64  \\
        \textbf{Inc-Res Block}   &2$\times$512 &512 &256 &128 \\
        \textbf{FC layer}       &1024          &None   &None   &None   \\
        \textbf{FC layer}       &10            &10  &10  &10  \\
        \hline        
        \textbf{Parameters (M)} &9.6  &3.2  &0.8 &0.2 \\
       \hline 
    \end{tabular}
    }
    \label{table:channel_setting} 
\end{table}

To further evaluate whether a wider or deeper neural network architecture is effective for ASC with the issue of mismatched recording devices, we replace the proposed CNN-based backbone by different benchmark network architectures of VGG16, VGG19, MobileNetV1, MobileNetV2, ResNet50V2, ResNet101V2, ResNet152V2, DenseNet121, DenseNet169, DenseNet201, Xception which are available from Keras Application API~\cite{keras_app}.
To this end, only the layers before the global pooling layer of these benchmark networks are used. 
These reused layers are then connected with the proposed MLP-based classification of the ASC baseline to perform end-to-end network architectures.
These network architectures are then evaluated and compared with the proposed ASC baseline.
Note that steps of the front-end feature extraction and the online data augmentation are remained during evaluating these network architectures.

\section{Ensemble methods and channel reduction for improving ASC performance} 
\label{baseline}

\subsection{Ensemble methods to improve accuracy} 
\label{ensemble}

As mentioned in Section~\ref{intro}, an ensemble of different input spectrograms or multiple models is a rule of thumb to enhance an ASC system performance. 
We, therefore, also evaluate these two ensemble strategies in this paper.
Regarding the ensemble of multiple spectrograms, we use three spectrograms of log-Mel, Constant Q Transform (CQT)~\cite{librosa_tool}, and Gammatone filter (Gam)~\cite{gam_filter}. 
By using the same settings mentioned in Section~\ref{feature}, all spectrograms present the same size of 128$\times$305$\times$3.
For each type of spectrogram, we apply the same data augmentation methods mentioned in Section~\ref{augmentation} and the proposed baseline presented in Section~\ref{network} for classification, referred to as CQT-baseline, log-Mel-baseline, Gam-baseline.
We then fuse the probability results by using PROD late fusion. 
In particular, we conduct experiments over individual network with different spectrogram inputs, then obtain predicted probability of each network as  \(\mathbf{\bar{p_{s}}}= (\bar{p}_{s1}, \bar{p}_{s2}, ..., \bar{p}_{sM})\), where $M$ is the category number and the \(s^{th}\) out of \(S\) networks evaluated. 
Next, the predicted probability after PROD fusion \(\mathbf{p_{prod}} = (\bar{p}_{1}, \bar{p}_{2}, ..., \bar{p}_{M}) \) is obtained by:
\begin{equation}
\label{eq:mix_up_x1}
\bar{p_{m}} = \frac{1}{S} \prod_{s=1}^{S} \bar{p}_{sm} ~~~  for  ~~ 1 \leq s \leq S 
\end{equation}
Finally, the predicted label  \(\hat{y}\) is determined by 
\begin{equation}
    \label{eq:label_determine}
    \hat{y} = arg max (\bar{p}_{1}, \bar{p}_{2}, ...,\bar{p}_{M} )
\end{equation}

In the second ensemble strategy, we evaluate whether ASC system performance can be improved by exploring information from sound events occurring in a sound scene recording.
In particular, we make use of the pre-trained CNN14 model from~\cite{kong_pretrain} which was trained with the large-scale AudioSet dataset~\cite{audioset} of 527 daily audio events.
We than feed log-Mel spectrograms into the pre-trained CNN14 model to extract sound-event-based embeddings which contains distinct features of sound events.
An embedding, which is a 2048-dimensional vector presenting for one input log-Mel spectrogram, is the feature map output of the global pooling layer of the pre-trained CNN14 model.
The embeddings are then fed into the proposed MLP-based classification of the ASC baseline without using Pooling Block for classifying into 10 categories, referred to as the sound-event-CNN14 system (SE-CNN14).
The predicted probability obtained from the sound-event-CNN14 system is than fused with results from CQT-baseline, log-Mel-baseline, and Gam-baseline using PROD fusion recently mentioned.    

\subsection{Channel reduction to achieve low complexity model} 
\label{channel_reduce}
To deal with the issue of large footprint model when using ensemble of multiple spectrograms, we apply the channel reduction technique.
In particular, the channel numbers at convolutional layers are reduced as shown in Table~\ref{table:channel_setting}.
We evaluate three cases of channel reduction, referred to as Red01, Red02, and Red03, which helps to reduce the model complexity of the proposed baseline from 9.6M, to 3.2M, 0.8M, and 0.2M of trainable parameters, respectively.

\section{Experiments and Discussion}

\subsection{Dataset and Metric}
\textbf{DCASE 2020 Task 1A Development set}~\cite{dcase2020_data}:The dataset comprises 23040 10-second segments with a total recording time of 64 hours. 
The dataset was recorded from three real devices namely A, B, and C with 40 hours, 3 hours, and 3 hours, respectively.
Additionally, synthesized audio recordings namely from S1 to S6 with 3-hour recording time for each are added.
As audio recordings are from both real and synthesized devices, this dataset is ideal to evaluate ASC task with the issue of mismatched recording devices.

We follow DCASE challenges, then separate the DCASE 2020 Task 1A Development set into Training and Evaluating subsets for training and evaluating processes, respectively (Note that audio recordings from S4, S5, and S6 are not presented in Training subset to evaluate unseen samples).
We also obey DCASE challenges, then use Accuracy (Acc.\%) as the metric for evaluating our proposed systems in this paper.
\subsection{Implementation}
As using the Mixup data augmentation method, labels are not one-hot encoding format. Therefore, we use Kullback–Leibler divergence (KL) loss shown in Eq. (\ref{eq:kl_loss}) below.
\begin{align}
   \label{eq:kl_loss}
   Loss_{KL}(\Theta) = \sum_{n=1}^{N}\mathbf{y}_{n}\log \left\{ \frac{\mathbf{y}_{n}}{\mathbf{\hat{y}}_{n}} \right\}  +  \frac{\lambda}{2}||\Theta||_{2}^{2}
\end{align}
where  \(\Theta\) are trainable parameters, constant \(\lambda\) is set initially to $0.0001$, $N$ is batch size set to 100, $\mathbf{y_{i}}$ and $\mathbf{\hat{y}_{i}}$  denote expected and predicted results.
We construct and train deep learning networks proposed with Tensorflow. 
We set epoch number=100 and using Adam method~\cite{Adam} for optimization.
While a learning rate of 0.0001 is set for the first 80 epoches with data augmentation methods, a low learning rate of 0.000001 is set for the next 20 epoches without any data augmentation method.
\subsection{Performance comparison among the proposed baseline and the benchmark network architectures}

As experimental results on DCASE 2020 Task 1A are shown in Table~\ref{table:res_01}, our proposed baseline system outperforms benchmark network architectures across recording devices.
Further analyse performance of benchmark network architectures, we can see that deeper neural networks such as VGG19, ResNet152V2 or DenseNet201 present low performance than the lower complexity networks such as VGG16, ResNet50V2, or DenseNet121 from the same architecture groups. 
These prove that a wider and shallow neural network is more effective rather than a deeper architecture for ASC task with mismatched recording devices.
\begin{table}[t]
    \caption{Performance comparison (Acc.\%) among log-Mel/CQT/Gam-baselines, sound-event-CNN14 system, and DCASE baseline on DCASE 2020 Task 1A Development set} 
        	\vspace{-0.2cm}
    \centering
    \scalebox{0.8}{

    \begin{tabular}{|c|c |c |c| c |c| } 
        \hline 
	         &\textbf{DCASE} & \textbf{log-Mel} &\textbf{Gam}  &\textbf{CQT}  &\textbf{Sound-event-}  \\
	         &\textbf{baseline} & \textbf{baseline} &\textbf{baseline}  &\textbf{baseline}  &\textbf{ CNN14 system}  \\

        \hline 
       
        \textbf{A(\%)}   &70.6 &77.3     &74.3  &61.5 &64.5\\
        \textbf{B(\%)}   &60.6 &72.0     &68.1  &55.9 &58.4\\
        \textbf{C(\%)}   &62.6 &76.6     &69.9  &53.2 &69.0\\
        \textbf{S1(\%)}  &55.0 &68.5     &64.8  &56.7 &51.8\\
        \textbf{S2(\%)}  &53.3 &65.8     &63.3  &49.7 &54.5\\
        \textbf{S3(\%)}  &51.7 &69.7     &66.7  &57.6 &56.4\\
        \textbf{S4(\%)}  &48.2 &63.3     &61.2  &51.2 &56.1\\
        \textbf{S5(\%)}  &45.2 &64.5     &63.0  &54.2 &55.5\\
        \textbf{S6(\%)}  &39.6 &58.8     &57.3  &47.0 &54.9\\
        \hline 
        \textbf{Average(\%)}  &54.1 &68.5 &65.4  &54.1 &57.9\\  
       \hline 
    \end{tabular}
    }
    \vspace{-0.3cm}
    \label{table:res_02} 
\end{table}
\begin{table}[t]
    \caption{Compare our proposed systems to 10 best systems from DCASE 2020 Task 1A challenge on Development set} 
        	\vspace{-0.2cm}
    \centering
    \scalebox{0.9}{

    \begin{tabular}{|c |c || c |c| } 
        \hline 
	         \textbf{Single Model} &\textbf{Acc.\%}  &\textbf{Ensemble}  &\textbf{Acc.\%}  \\
        \hline 
             Top-1    &72.1       &Top-1       &84.2  \\
             Top-2    &68.9       &Top-2       &75.0  \\
             Top-3    &73.7       &Top-3       &74.4  \\
             Top-4    &71.8       &Top-4       &73.3  \\
             Top-5    &70.2       &Top-5       &73.1  \\
             Top-6    &72.7       &Top-6       &-  \\
             Top-7    &71.8       &Top-7       &72.5  \\
             Top-8    &72.1       &Top-8       &-  \\
             Top-9    &68.9       &Top-9       &71.9  \\
             Top-10   &-          &Top-10      &71.9  \\
        \hline 
              \textbf{Our log-Mel-baseline}   &68.5       &\textbf{Our ensemble}      &74.7  \\
       \hline 
    \end{tabular}
    }
    \vspace{-0.4cm}
    \label{table:res_03} 
\end{table}
\subsection{Performance comparison among spectrograms and sound-event-based systems}

As Table~\ref{table:res_02} shows, while log-Mel and Gam spectrograms are competitive, CQT shows lower performance.
As our proposed baseline, log-Mel baseline, significantly outperform the sound-event-based systems (DCASE baseline with the pre-trained OpenL3~\cite{openl3_model} or the proposed sound-event-CNN14 system with the pre-trained CNN14~\cite{kong_pretrain}) over all devices, it can be concluded that directly training a model on sound scene recordings is more effective than systems basing on  sound-event-based embeddings.

\subsection{How ensembles and channel reduction help to improve ASC performance}
Table~\ref{table:res_04} shows how ensemble methods and channel reduction help to improve ASC performance.
Although applying channel reduction technique helps to reduce the model complexity, it reduces the accuracy performance on almost single models, excepted CQT with Red02.
By combining ensembles of multiple spectrograms and channel reduction, we can achieve good models which show a balance between accuracy performance and complexity (i.e. Ensembles of three spectrograms with Red01 and Red02 present 70.6\% with 9.6M and 69.9\% with 2.4M, respectively).
When the result of sound-event-event system, SE-CNN14, is fused, this helps to further improve the performance by an average of 2.5\% as shown in the bottom part of Table~\ref{table:res_04}. 
The best accuracy of 74.7\% is obtained from an ensemble of the best single models: CQT w/ Red02, Gam-baseline, log-Mel-baseline, and SE-CNN14.
However, as SE-CNN14 presents a large footprint of 80.7M, it makes ensemble models significantly complex.
%
\begin{table}[t]
    \caption{Performance comparison among single and ensemble models with or without channel reduction on DCASE 2020 Task 1A Development set} 
        	\vspace{-0.2cm}
    \centering
    \scalebox{0.85}{

    \begin{tabular}{|c|c |c | } 
        \hline        
	    \textbf{Single Models}   &\textbf{Acc.(\%)} &\textbf{Parameters (M)}\\
        \hline        
        CQT-baseline   &54.1 &9.6\\
        CQT w/ Red01      &56.8 &3.2\\
        CQT w/ Red02      &\textbf{58.0} &0.8\\
        CQT w/ Red03      &55.1 &0.2\\
            \hline                                                   
        Gam-baseline   &\textbf{65.4} &9.6\\
        Gam w/ Red01      &62.1 &3.2\\
        Gam w/ Red02      &58.5 &0.8\\
        Gam w/ Red03      &58.0 &0.2\\
            \hline                                                   
        log-Mel-baseline   &\textbf{68.5} &9.6\\
        log-Mel w/ Red01      &65.0 &3.2\\
        log-Mel w/ Red02      &62.0 &0.8\\
        log-Mel w/ Red03      &60.0 &0.2\\
            \hline        
	\textbf{Ensemble Models}   &\textbf{Acc.(\%)} &\textbf{Parameters (M)}\\
            \hline                                                   
        3 baselines   &\textbf{71.1} &28.8\\
        3 baselines w/ Red01      &70.6 &9.6\\
        3 baselines w/ Red02      &69.9 &2.4\\
        3 baselines w/ Red03      &67.9 &0.6\\
            \hline                                                   
        3 baselines + SE-CNN14   &\textbf{73.8} &109.5\\
        3 baselines w/ Red01 + SE-CNN14      &72.8 &90.3\\
        3 baselines w/ Red02 + SE-CNN14      &72.4 &83.1\\
        3 baselines w/ Red03 + SE-CNN14      &71.0 &81.3\\
       \hline 
    \end{tabular}
    }
    \vspace{-0.5cm}
    \label{table:res_04} 
\end{table}
%
\subsection{Compare with the state-of-the-art systems}
Table~\ref{table:res_03} compares our best single model (log-Mel-baseline) and ensemble model with ten best systems submitted to DCASE 2020 Task 1A challenge~\cite{dcase_2020_1a}.
Our proposed single model, log-Mel-baseline, not only shows competitive performance to the state-of-the-art ASC systems, but it also presents a very shallow network architecture (i.e. While almost single model in Table~\ref{table:res_03} mainly use deep residual-based architectures, the log-Mel-baseline presents a deepness of 8-convolutional layers and two-fully-connected layers).
Regarding ensemble models, our best system (CQT w/ Red02, Gam-baseline, log-Mel-baseline, and SE-CNN14) achieves the top-3 ranking which records an accuracy of 74.7\%.
A variant model, which also makes use the ensemble of multiple spectrograms, the shallow network architecture, and channel reduction, was also submitted to DCASE 2021 Task 1A (i.e. This challenge requires a very low footprint model with less than 32.7K of trainable parameters) and achieved the top-9 team ranking~\cite{dcase_2021_1a}. 

\section{Conclusion}
This paper has presented a novel inception-residual-based neural network for ASC task with mismatched recording devices.
By conducting intensive experiments over the benchmark DCASE 2020 Task 1A Development dataset, it is indicated that the novel network presenting a wider and shallow architecture is more effective for ASC rather than deeper architectures. 
Additionally, our proposed ensemble of multiple spectrograms and channel reduction (e.g. Red02) help to achieve an accuracy of 69.9\% and low footprint of 2.4M trainable parameters, which shows an balance between performance and model complexity.
These results also prove that our proposed system is competitive to the state-of-the-art systems and validates ASC application on edge devices.

\section{Acknowledgements}

The AMMONIS project is funded by the FORTE program of the Austrian Research Promotion Agency (FFG) and the  Federal Ministry of Agriculture, Regions and Tourism (BMLRT) under grant no. 879705.

\bibliographystyle{IEEEtran}

\bibliography{mybib}


\end{document}